\PassOptionsToPackage{unicode}{hyperref}
\PassOptionsToPackage{hyphens}{url}
\PassOptionsToPackage{dvipsnames,svgnames,x11names}{xcolor}
\documentclass[
  letterpaper,
  DIV=11,
  numbers=noendperiod]{scrartcl}
\usepackage{xcolor}
\usepackage{amsmath,amssymb}
\usepackage{iftex}
\ifPDFTeX
  \usepackage[T1]{fontenc}
  \usepackage[utf8]{inputenc}
  \usepackage{textcomp} 
\else 
  \usepackage{unicode-math} 
  \defaultfontfeatures{Scale=MatchLowercase}
  \defaultfontfeatures[\rmfamily]{Ligatures=TeX,Scale=1}
\fi
\usepackage{lmodern}
\ifPDFTeX\else
\fi
\IfFileExists{upquote.sty}{\usepackage{upquote}}{}
\IfFileExists{microtype.sty}{
  \usepackage[]{microtype}
  \UseMicrotypeSet[protrusion]{basicmath} 
}{}
\makeatletter
\@ifundefined{KOMAClassName}{
  \IfFileExists{parskip.sty}{%
    \usepackage{parskip}
  }{
    \setlength{\parindent}{0pt}
    \setlength{\parskip}{6pt plus 2pt minus 1pt}}
}{
  \KOMAoptions{parskip=half}}
\makeatother
\makeatletter
\ifx\paragraph\undefined\else
  \let\oldparagraph\paragraph
  \renewcommand{\paragraph}{
    \@ifstar
      \xxxParagraphStar
      \xxxParagraphNoStar
  }
  \newcommand{\xxxParagraphStar}[1]{\oldparagraph*{#1}\mbox{}}
  \newcommand{\xxxParagraphNoStar}[1]{\oldparagraph{#1}\mbox{}}
\fi
\ifx\subparagraph\undefined\else
  \let\oldsubparagraph\subparagraph
  \renewcommand{\subparagraph}{
    \@ifstar
      \xxxSubParagraphStar
      \xxxSubParagraphNoStar
  }
  \newcommand{\xxxSubParagraphStar}[1]{\oldsubparagraph*{#1}\mbox{}}
  \newcommand{\xxxSubParagraphNoStar}[1]{\oldsubparagraph{#1}\mbox{}}
\fi
\makeatother

\usepackage{longtable,booktabs,array}
\usepackage{calc} 
\usepackage{etoolbox}
\makeatletter
\patchcmd\longtable{\par}{\if@noskipsec\mbox{}\fi\par}{}{}
\makeatother
\IfFileExists{footnotehyper.sty}{\usepackage{footnotehyper}}{\usepackage{footnote}}
\makesavenoteenv{longtable}
\usepackage{graphicx}
\makeatletter
\newsavebox\pandoc@box
\newcommand*\pandocbounded[1]{
  \sbox\pandoc@box{#1}%
  \Gscale@div\@tempa{\textheight}{\dimexpr\ht\pandoc@box+\dp\pandoc@box\relax}%
  \Gscale@div\@tempb{\linewidth}{\wd\pandoc@box}%
  \ifdim\@tempb\p@<\@tempa\p@\let\@tempa\@tempb\fi
  \ifdim\@tempa\p@<\p@\scalebox{\@tempa}{\usebox\pandoc@box}%
  \else\usebox{\pandoc@box}%
  \fi%
}
\def\fps@figure{htbp}
\makeatother

\NewDocumentCommand\citeproctext{}{}

\makeatletter
 \let\@cite@ofmt\@firstofone
 \def\@biblabel#1{}
 \def\@cite#1#2{{#1\if@tempswa , #2\fi}}
\makeatother
\newlength{\cslhangindent}
\newlength{\csllabelwidth}
\newenvironment{CSLReferences}[2] 
 {\begin{list}{}{%
  \setlength{\itemindent}{0pt}
  \setlength{\leftmargin}{0pt}
  \setlength{\parsep}{0pt}
  \ifodd #1
   \setlength{\leftmargin}{\cslhangindent}
   \setlength{\itemindent}{-1\cslhangindent}
  \fi
  \setlength{\itemsep}{#2\baselineskip}}}
 {\end{list}}
\usepackage{calc}

\providecommand{\tightlist}{%
  \setlength{\itemsep}{0pt}\setlength{\parskip}{0pt}}

\KOMAoption{captions}{tableheading}
\makeatletter
\@ifpackageloaded{caption}{}{\usepackage{caption}}
\AtBeginDocument{%
\ifdefined\contentsname
  \renewcommand*\contentsname{Table of contents}
\else
  \newcommand\contentsname{Table of contents}
\fi
\ifdefined\listfigurename
  \renewcommand*\listfigurename{List of Figures}
\else
  \newcommand\listfigurename{List of Figures}
\fi
\ifdefined\listtablename
  \renewcommand*\listtablename{List of Tables}
\else
  \newcommand\listtablename{List of Tables}
\fi
\ifdefined\figurename
  \renewcommand*\figurename{Figure}
\else
  \newcommand\figurename{Figure}
\fi
\ifdefined\tablename
  \renewcommand*\tablename{Table}
\else
  \newcommand\tablename{Table}
\fi
}
\@ifpackageloaded{float}{}{\usepackage{float}}
\floatstyle{ruled}
\@ifundefined{c@chapter}{\newfloat{codelisting}{h}{lop}}{\newfloat{codelisting}{h}{lop}[chapter]}
\floatname{codelisting}{Listing}

\makeatother
\makeatletter
\@ifpackageloaded{caption}{}{\usepackage{caption}}
\@ifpackageloaded{subcaption}{}{\usepackage{subcaption}}
\makeatother
\usepackage{bookmark}
\IfFileExists{xurl.sty}{\usepackage{xurl}}{} 
\hypersetup{
  pdftitle={Synthetic Data and the Shifting Ground of Truth},
  pdfauthor={Dietmar Offenhuber},
  colorlinks=true,
  linkcolor={blue},
  filecolor={Maroon},
  citecolor={Blue},
  urlcolor={Blue},
  pdfcreator={LaTeX via pandoc}}

\title{Synthetic Data and the Shifting Ground of Truth}
\usepackage{etoolbox}
\makeatletter
\providecommand{\subtitle}[1]{
  \apptocmd{\@title}{\par {\large #1 \par}}{}{}
}
\makeatother
\subtitle{Talk presented at the Society for the Social Studies of
Science (4S) 2025 meeting in Seattle.}
\author{Dietmar Offenhuber}
\date{2025-09-03}
\begin{document}
\maketitle

\subsection{Abstract}\label{abstract}

The emergence of synthetic data for privacy protection, training data
generation, or simply convenient access to quasi-realistic data in any
shape or volume complicates the concept of ground truth. Synthetic data
mimic real-world observations, but do not refer to external features.
This lack of a representational relationship, however, not prevent
researchers from using synthetic data as training data for AI models and
ground truth repositories. It is claimed that the lack of data realism
is not merely an acceptable tradeoff, but often leads to better model
performance than realistic data: compensate for known biases, prevent
overfitting and support generalization, and make the models more robust
in dealing with unexpected outliers. Indeed, injecting noisy and
outright implausible data into training sets can be beneficial for the
model. This greatly complicates usual assumptions based on which
representational accuracy determines data fidelity (garbage in - garbage
out). Furthermore, ground truth becomes a self-referential affair, in
which the labels used as a ground truth repository are themselves
synthetic products of a generative model and as such not connected to
real-world observations. My paper examines how ML researchers and
practitioners bootstrap ground truth under such paradoxical
circumstances without relying on the stable ground of representation and
real-world reference. It will also reflect on the broader implications
of a shift from a representational to what could be described as a
mimetic or iconic concept of data.

\begin{figure}[H]

{\centering \pandocbounded{\includegraphics[keepaspectratio]{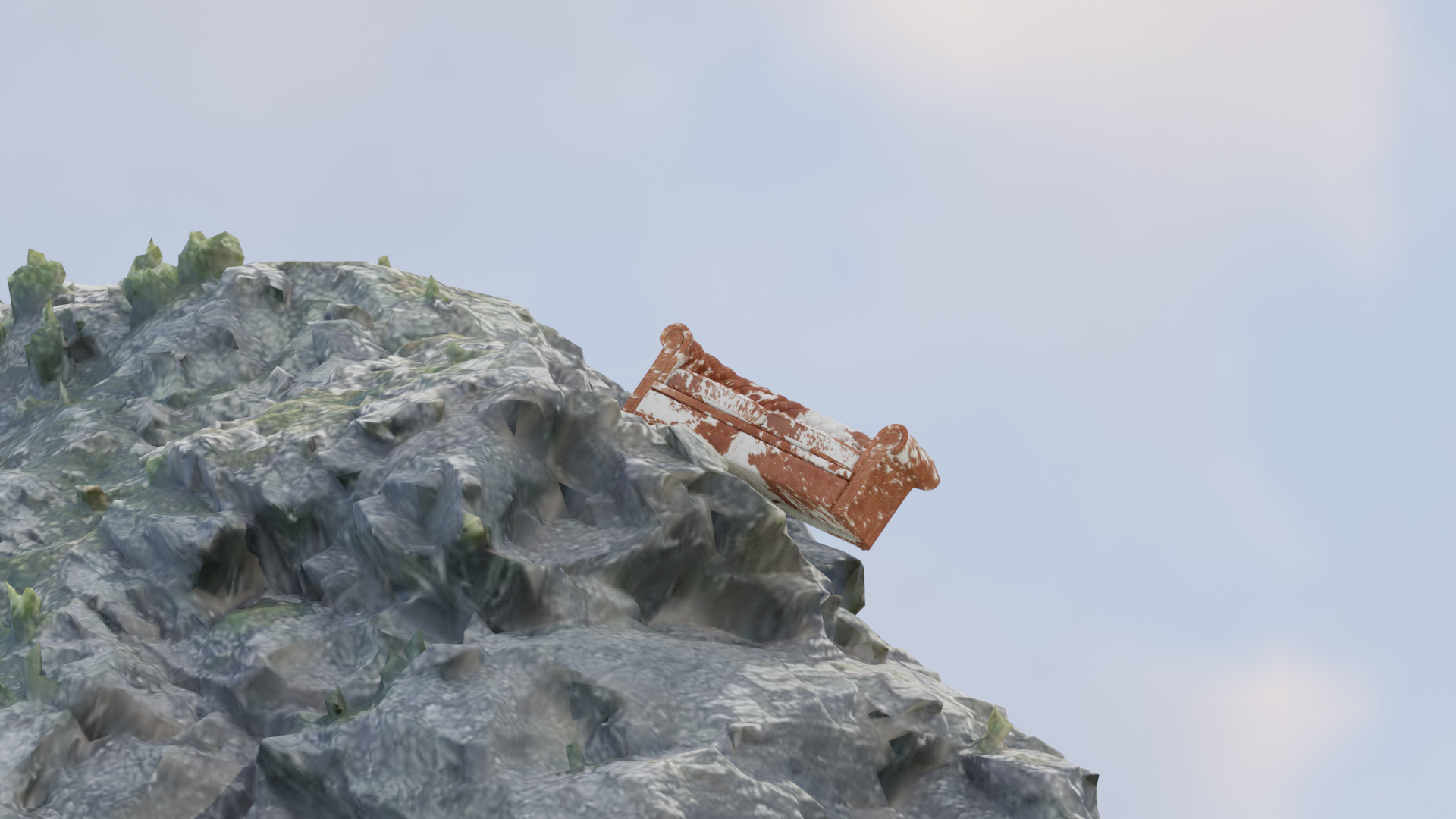}}

}

\caption{Simone Niquille, Still from ``Sorting Song'' 2021}

\end{figure}%

\subsection{Introduction}\label{introduction}

The concept of ground truth, traditionally serving as a stable reference
point for evaluating model predictions, undergoes significant
transformation when applied to synthetic data. This paper examines the
epistemic and ontological foundations of the concept of ground truth,
used differently in different contexts. Gil Fournier and Jussi Parikka,
describe ground truth broadly as a ``grounding figure of knowledge,''
both a guarantee of knowledge and a metaphorical figure (Gil-Fournier
and Parikka 2020). In the case of synthetic data, however, this
grounding figure undergoes several figure-ground inversions: what was
once considered the static background, turns into a site of uncertainty
and controversy. These shifts turn the common representational
understanding of data on its head.

To begin, consider the following set of conflicting assumptions:

\textbf{Garbage in - garbage out}: A common stance of critical data
studies is that every machine learning model is only as good as its
training data. Good in this context means how well and complete it
represents its population or phenomenon. It has been closely studied how
biased training data automates and perpetuates historical inequalities
and exclusion (Benjamin 2019; D'Ignazio and Klein 2020)

\textbf{Counter-intuitive findings in AI research}: Seemingly in
conflict with this stance, AI resarch has shown that including fake,
improbable, even impossible observations into training data can improve
learning and generalization; make models more robust. As Tobin et
al.~note, ``With enough variability in the simulator, the real world may
appear to the model as just another variation'' (Tobin et al. 2017).
\emph{Increasing variability} in this context means precisely not to
make synthetic data as realistic as possible, but to explore the full
\emph{parameter space} to force a machine learning model to generalize
and focus on the big picture rather than smaller, often spurious
correlations.

My paper will discuss the consequences of this paradoxon for the concept
of ground truth.

\subsection{A brief typology of synthetic
data}\label{a-brief-typology-of-synthetic-data}

It is important to avoid overly-general statements about synthetic data,
as this concept is an umbrella term for data generated for very
different purposes with very different functions and properties.
Broadly, it means any data that is artificially generated for a
particular purpose. What follows are the four most important, as
discussed in Offenhuber (2024).

\begin{enumerate}
\def\labelenumi{\arabic{enumi}.}
\tightlist
\item
  \textbf{Algorithmic} - this includes the algorithmically generated
  data sets and simulations that are used as benchmarks in disciplines
  such as physics or network science. synthetic data here is a idealized
  case that real-world data can be compared to.
\item
  \textbf{Obfuscated} - real observations, obfuscated for the purpose of
  privacy protection (i.e.~of micro-census data, which uses noise
  injection for differential privacy). mathematical theory around
  differential privacy that allows to calculate how much obfuscation is
  needed.
\item
  \textbf{Mimikry, hybridized} - data generated based on one or several
  data sets, for a variety of reasons (analytic purposes with issues of
  privacy, de-biasing, test data). Involves combining different datasets
  to increase resolution, or enlarge a data set preserving known
  statistical properties (not necessarily the unknown ones). also
  possible to Photoshop data that way, reduce known biases etc\ldots{}
  but the important thing is that here we no longer talk about real
  observations, but data that look like real observations.
\item
  \textbf{Generated training data} - entirely artificial data sets for
  training purposes that are not generated based on real data,
  completely invented. here we are very far away from any
  representational relationship or fidelity. the only objective is that
  these data improve learning when injected into other observations.
\end{enumerate}

The following discussion is not equally relevant for all four
categories; it mostly applies to the last two, although some arguments
also apply to second category focusing on differential privacy
obfuscation.

\subsection{The rationales for synthetic
data}\label{the-rationales-for-synthetic-data}

There is common ground between the garbage-in, garbage-out stance and
the case for increasing variability through generalization. In both
cases, the core problem is well understood: all observational training
data are biased and contain stereotypes. They also don't cover enough
outliers. Inclusion of outliers is considered important in both cases,
although for slightly different reasons. In the first case, the goal is
proportionally accurate representation. The second case aims to cover
all possible variations, regardless of proportionality. Another goal is
to increase the contrast between features to put smaller variations into
perspective.

Microsoft's \emph{Face Synthetics} dataset is designed to train facial
recognition models without using photos of real people. Instead, the
dataset contains rendered 3D characters with randomized features, facial
expressions, and lighting conditions (Castellanos 2021). While this
randomization may not accurately reflect demographic variations, it
attempts to maximize diversity. After all, a model should have as much
training data relating to underrepresented cases as to those that are
more common.

AI research on underrepresented cases is limited by a lack of real-world
training data, which puts models at risk of underperforming for these
populations. In the medical field, rare brain tumors are outliers.
Researchers have replaced real MRI images with synthetically generated
images of brain tumors and subsequently used them to train a model to
recognize real tumors (Rouzrokh et al., n.d.). As in the facial
recognition example, the goal is a simulation-to-reality transfer. In
this case, however, it is complicated by the scarcity of real-world
medical data on rare conditions.

An even more extreme case are data about phenomena that have never been
observed or events that are not known to have ever occurred. In this
context, Jacobsen addresses the political dimension of synthetic data,
driven from the desire to model ``black swan events'' to manage risks in
technocratic regimes of governance (Jacobsen 2023). Steinhoff also
considers synthetic data to follow but eventually supersede the paradigm
of surveillance capitalism, While surveillance is expensive, slow, and
labor intensive, synthetic data is cheap, but may just be as effective
for applications such as consumer behavior modeling (Steinhoff 2022).

\subsection{Figure-ground inversions of synthetic
data}\label{figure-ground-inversions-of-synthetic-data}

How does synthetic data fit into familiar concepts of data and their
epistemic functions? I think there is quite a lot.

\subsubsection{1. Inversion - from representation to
imitation}\label{inversion---from-representation-to-imitation}

The concept of data as representations of reality is inherently
problematic (Coopmans, Vertesi, and Lynch 2014). Critics point out that
data receive meaning not only based on their content, they are not mere
descriptions. Instead data receive meaning through the context of social
practices. Furthermore, data and the world co-constitute each other,
while representation considers the world as static (Kitchin and Dodge
2007).

An alternative relational concept of data foregrounds data as material
artifacts used in social practices - ``objects that are explicitly
collected and disseminated in order to provide evidence for claims about
reality'' (Leonelli 2016; see also Latour 1990). Rather than considering
the world as a stable referent, data and the world continuously
co-constitute each other e.g.~through their use in discourse. However,
even these critiques do not challenge that data often still has an
important representational function.

Synthetic data takes forces us to take these critiques further, because
now we are confronted with data that does not involve any
representational relationships. Instead, it somehow behaves like real
data, substitutes and augments in a specific application context. Rather
than serving as symbolic or indexical references, synthetic data are
characterized by an iconic or mimetic relationship to the
object.\footnote{We can still speak of an indexical relationship with
  the model that generated the synthetic data set - this is an argument
  made in my previous paper Offenhuber (2024).}

The sometimes-absurd consequences of this shift become visible in the
work of the artist Simone C Niquille.\footnote{See her website
  \url{https://www.technofle.sh}} She works with various synthetic
training datasets used for machine vision systems. In her critical
practice, she takes their representational function literally, examining
how people, living spaces, and cities are represented within these
datasets.

\begin{figure}[H]

{\centering \pandocbounded{\includegraphics[keepaspectratio]{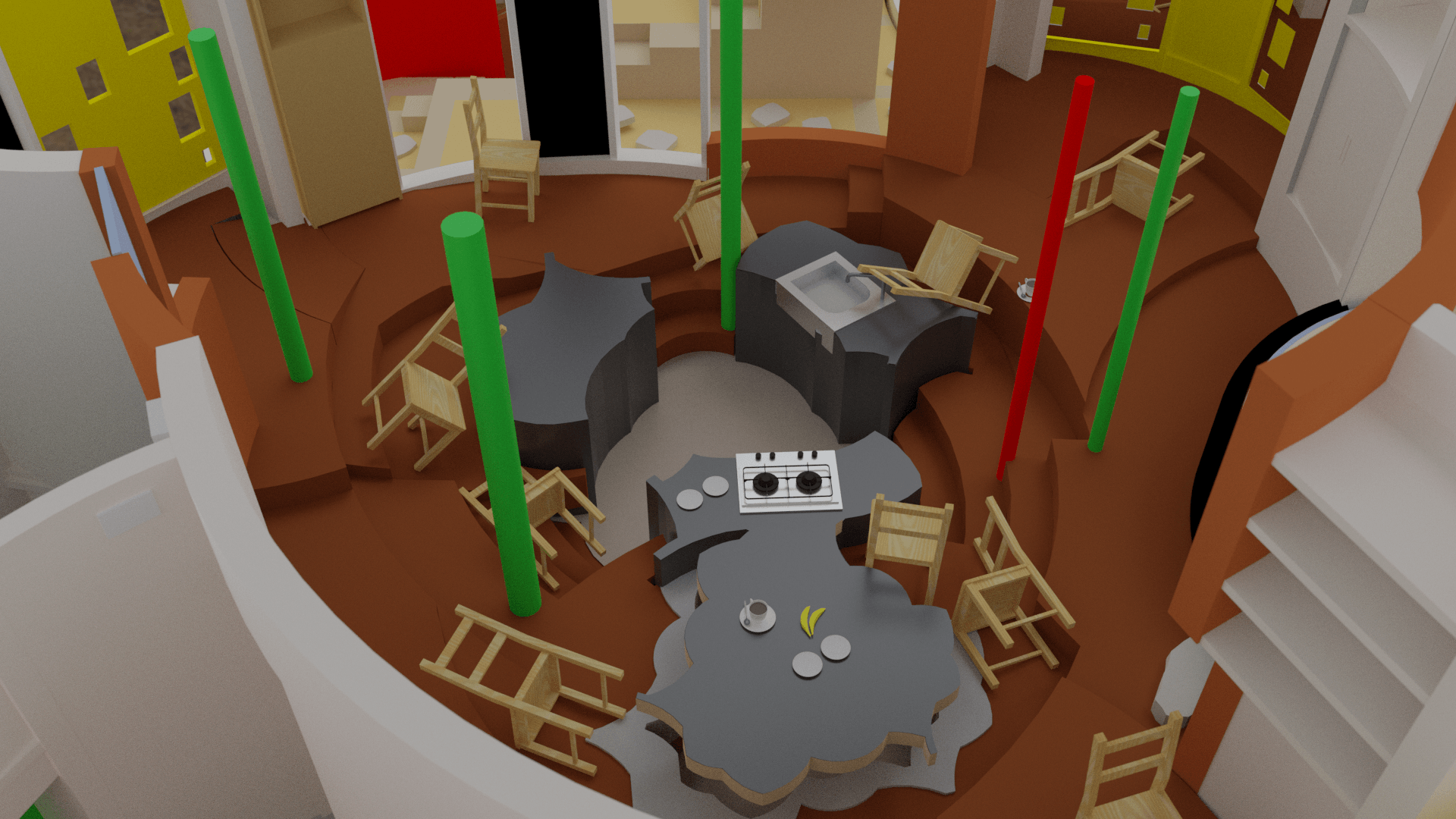}}

}

\caption{Dawn, all is asleep. The apartment humming of charging
batteries and the faint beep of a completed cleaning cycle. Simone
Niquille, in collaboration with ST Luk of the Reversible Destiny
Foundation, IKEA Bertil chair at the Arakawa + Gins Shidami apartment
(2021) © Simone Niquille / Technoflesh (2021). Shidami apartment project
© Arakawa + Gins}

\end{figure}%

This shift from an representational to an mimetic function has some
interesting consequences:

We no longer have a straightforward way to establish data quality,
e.g.~by measuring how well a data set represents a population, even when
considering multiple perspectives and ambiguities. The quality of a
synthetic dataset, on the other hand, depends on the goal: how well it
preserves privacy in a particular application, or the quality of
predictions of a model trained with it.

This means that data quality can no longer be considered in relation to
reality (whether shared or pluriversal), but only in relation to a
particular purpose. For example, the degree to which census data is
obscured for privacy purposes may be appropriate for one use case but
not another. In other words, data quality is no longer about what the
data \emph{represent}, but rather, what the data \emph{do}.

\subsubsection{2. Inversion of provenance: from evidence to training
signal}\label{inversion-of-provenance-from-evidence-to-training-signal}

Data quality that cannot be determined \emph{a priori}, but only after
use in a particular scenario leads to another figure-ground inversion.
For data to function as evidence requires demonstrating a chain of
custody back to their origin; to justify their accuracy and reliability.
The reference from data to its object needs to be traceable.

In the case of synthetic data, the arrow is reversed, pointing forward
to the performance of the model trained for a particular situation. Data
sets are constructed based on the desired result. If computer vision
model is biased, because it does not recognize people wearing hats, one
has to paint hats onto training images. If there are not enough MRIs
with a certain brain tumor, one has to generate MRIs featuring tumors
with an generative AI model, and one can then measure how well the model
then recognizes real tumors.

So instead of a correspondence theory of truth, which, again, is
problematic in itself, we have a teleological theory of truth, which is
a contradiction in itself. A better way to say it is perhaps that the
correspondence requirement has shifted from data to the trained model,
which moves into the foreground. The behavior of this model, especially
with regard to outliers, is not easily testable or traceable: instead of
a relatively simple representational correspondence, we now have to
consider a complex set of metrics.

\subsubsection{3. The concept of ground truth is
shifting}\label{the-concept-of-ground-truth-is-shifting}

These circumstances leads to several shifts in the concept of ground
truth. As a contribution to a recent complexity science workshop in
Vienna put it in their data physicalization project, the ground in
ground truth no longer refers to a stable reference, but is ground like
coffee, boiled and sent through a filter that selectively withholds
components.

\begin{figure}[H]

{\centering \pandocbounded{\includegraphics[keepaspectratio]{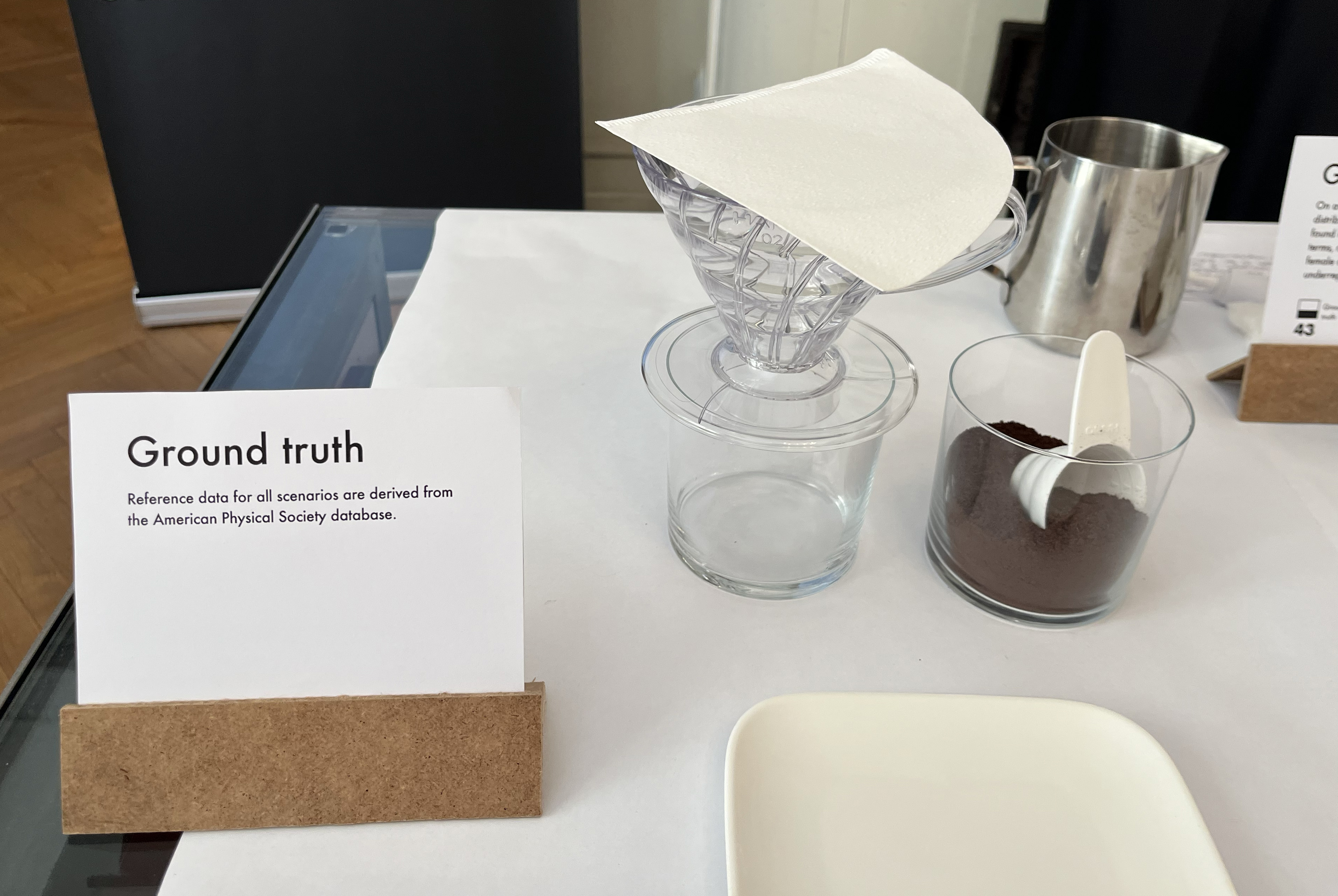}}

}

\caption{Ground Truth - project at the Visualization workshop at the
Vienna Complexity Science Hub, August 2025. By Kelly Krause, Jason
Dilworth, Katherine Poblete, Naomi Hinkelmann, Timo Schnepf}

\end{figure}%

To explore shifting notions of ground truth, I spoke with ML experts and
reviewed recent literature. I should mention on the outset that the term
ground truth is controversial even among ML researchers. Many of them
say that the term should not be used at all.

In the remaining sections, I will distinguish between three ways how the
term is used in recent literature:

\begin{enumerate}
\def\labelenumi{\arabic{enumi}.}
\tightlist
\item
  Ground truth as the best available measure of external validity
\item
  Ground truth as whatever labelers (human or automatic) label
\item
  Disagreement about truth as productive for training
\end{enumerate}

\paragraph{1. Ground truth as the best available measure of external
validity}\label{ground-truth-as-the-best-available-measure-of-external-validity}

Traditionally, ground truth refers to the best available information.
This ranges from positivist positions that assume reality to be shared
and stable to more contextual ones.

On the positivist end of the spectrum, we find the classic understanding
of ground truth in remote sensing and a military context, where it
refers to surface observation data (Hoffer 1972). Think of Trump's
statement
``\href{https://www.whitehouse.gov/articles/2025/06/irans-nuclear-facilities-have-been-obliterated-and-suggestions-otherwise-are-fake-news/}{Iran's
Nuclear Facilities Have Been Obliterated}'' such statements require
verification on the geographical surface, and someone has to go and
check. Of course, even this classic meaning is more complicated, as
ground truth is established in a relationship between geography and
images, that is ``heavily overdetermined'' by many forms of
representational biases (Gil-Fournier and Parikka 2020). Gil-Fournier
and Parikka describe ground truth as a grounding figure of knowledge,
critiquing the assumed factuality of geographic information by calling
attention to all the image operations involved in establishing these
facts.

A different, interesting twist is provided by Bjerre-Nielsen and Glavind
(2022) who describe the role of ethnographic observations as a possible
ground truth for big datasets (e.g.~exhaust), contextualizing and
complementing such datasets with a thick description. This is of course
now very different from a positivist notion of ground truth. This
perspective still tries to establish ``whether the data measures what
was assumed'', going even further, arguing that ethnographic fieldwork
supports causal inference, because the ethnographer can intervene in the
situation, and therefore introduce experimental variation.

Cabitza, Ciucci, and Rasoini (2019) describes ground truth as a
manifold, distinguishing between what computer scientists and physicians
pursue in medical data - for the CS folks, ground truth means accuracy
and completeness, for the physicians it is trustworthiness and
meaningfulness. For physicians, bias is less of a problem, because they
already consider every case to be somewhat unique.

Jaton (2017) finally writes about how ground truth in ML is already
structured based on the design goal, embedding epistemic assumptions.
Therefore, ``we get the algorithms of our ground truths'' that is,
algorithms are ``sets of instructions designed to computationally
retrieve in the best possible way what have been designed as outputs
during specific problematization processes.'' here, ground truth is
strictly problem-focused, the truth does not exist outside this problem
formulation. Ground truth is the sandbox designed for the algorithm,
which implicitly embodies the algorithmic logic. Here one may note a
resonance with the second inversion of synthetic data, the centering on
the problem context and model performance.

\paragraph{2. Ground truth as whatever labelers
label}\label{ground-truth-as-whatever-labelers-label}

In this opposed perspective, the concept of external truth is no longer
considered central. Ground truth can be observational data, but does not
have to be. This perspective focuses less on ontological claims, and
instead uses ground truth as a strictly operational definition. It may
simply also accept that ground truth is unattainable ``Ground-truth
refers to the fact that researchers (at least those not working at
Twitter) cannot say with certainty how many fully or partially automated
accounts exist in the total population'' (Martini et al. 2021).

As Grosman et al.~explain, ``The ``ground truth'\,' part refers to the
categories or labels which humans---e.g.~domain experts, computer
scientists or Amazon turkers---have attributed to each sequence. It
supplies the system with answers to the problem: the algorithm now has
an external check for assessing the correctness of its classification''
(Grosman and Reigeluth 2019). This external reference, however, may be
unreliable, or its reliability unknown.

Labeled data may also be unreliable due to social distance, when
labeling is outsourced to low-wage countries. ``Even though the computer
vision model is designed in Paris by a French start-up, the data
annotation needed to build the ground truth datasets necessary to train
the model is delegated to a contractor in Madagascar'' (Le Ludec,
Cornet, and Casilli 2023).

Data may also be labeled by other AI models. In this approach, human
labels are compared to machine labeling in order to interpret
differences without making a judgment which one is considered more
reliable (Rieder and Skop 2021).

In the context of synthetic data, the ground truth is often entirely
AI‑generated. This raises an institutional‑political issue:
machine‑generated ground truth is deemed reliable largely because it
originates from a major tech firm such as Google, and is promoted as a
standardized reference. As McGuigan notes, ``By positioning Google's
algorithmic models as the ground truth on which publishers and
advertisers can evaluate the effectiveness of their
placements''(McGuigan et al. 2023).

Training data sets are automatically labeled by other models are often
too voluminous to be verified in detail. Again, we encounter our
figure-ground reversal. The researcher knows the intended behavior of
the model, yet has little insight into the ontological status or
reliability of the training data that underpins it.

\paragraph{3. Disagreements about truth are
productive}\label{disagreements-about-truth-are-productive}

This third group of papers rejects the notion of a single gold standard
and instead treats disagreements among annotators as a valuable training
information. As Aroyo et al.~argue, ``disagreement is not noise but
signal'' (Aroyo and Welty 2015). Devani et \,al.~show that such
disagreement can capture important nuances and reveal meaningful
ambiguities. Their multi‑annotator framework explicitly models annotator
uncertainty, which helps identify systematic biases and ``scenarios
where knowing when not to make a prediction is important'' (Davani,
Díaz, and Prabhakaran 2022).

Human uncertainty distributions can also be leveraged during training.
Peterson notes that ``errors in classification can be just as
informative as the correct answers---a network that confuses a dog with
a cat, for example, might be judged to generalize better than one that
confuses it with a truck''(Peterson et al. 2019). However, Uma et\,
al.~caution that multi‑annotator models may simply shift the problem
from data evaluation to model evaluation (Uma et al. 2022). They
emphasize that even without a gold standard, it remains essential to
agree on how to assess model performance. Thus, this line of work
operationalizes the earlier call for increased variability in training
data.

\subsubsection{Conclusion}\label{conclusion}

When representational correspondence and fidelity are no longer the
primary criteria for evaluating data, the task of evaluation itself
becomes unsettled. Synthetic data complicate the long-standing
association of data with representation, accuracy, or evidential
grounding. They are not ``things given'' in the courtroom sense, nor
even ``capta,'' in Johanna Drucker's formulation of ``things taken,''
which emphasizes the subjectivity of collection (Drucker 2011). Both of
these concepts still frame data as passive artifacts---whether
objectively recorded or subjectively extracted.

Synthetic data, by contrast, act as active agents. They shape the
learning of models in situated contexts without necessarily describing
or representing external states of affairs. Their value does not lie in
fidelity to an external world. In this sense, synthetic data mark a
departure from the evidentiary paradigm. They do not serve as guarantees
of truth but as instruments of variation and transformation. To take
synthetic data seriously thus requires rethinking data as
interventions---operations that generate, perturb, and reconstitute the
very grounds on which claims to knowledge are made.

\subsection*{References}\label{references}
\addcontentsline{toc}{subsection}{References}

\phantomsection\label{refs}

\end{document}